# *Ab initio* study of ferromagnetism induced by magnetic impurities in rutile $TiO_2$

**L. A. Errico**[*,1], **M. Weissmann**[2], and **M. Rentería**[1]

[1] Departamento de Física, Facultad de Ciencias Exactas, Universidad Nacional de La Plata, CC No 67, 1900 La Plata, Argentina
[2] Departamento de Física, Comisión Nacional de Energía Atómica, Avda. del Libertador 8250, 1429 Buenos Aires, Argentina

Using the first-principles density-functional approach, magnetic properties of Mn-, Fe-, Co-, and Ni-doped rutile $TiO_2$ were investigated for two different impurity concentrations (25% and 6.25%). Calculations were performed with the Full-Potential Linearized-Augmented Plane Waves (FLAPW) method, assuming that the magnetic impurities substitutionally replace the Ti ions. Our results show that the systems (with the exception of Ni-doped $TiO_2$) are ferromagnetic. We also found that polarization mainly occurs at the impurity sites, and the magnetic moments of the impurities are independent of the impurity concentration.

## 1 Introduction

Integrating spin functionality into otherwise nonmagnetic materials has become a highly desirable goal in the last years. In particular, dilute magnetic impurities in semiconductors and oxides produce novel materials appealing for spintronics and optoelectronics (see, e.g., Ref. [1] and references therein). Magnetic dopants in nonmagnetic solids are assumed to couple with the electronic states of the host, but remaining magnetically active. For their practical applications, ferromagnetic semiconductors are required to have a high Curie temperature ($T_C$). While most of the dilute magnetic semiconductors (like $Ga_{1-x}Mn_x$) have a $T_C$ much lower than room temperature, room-temperature ferromagnetism has been observed in Mn-doped compounds such as $CdGeP_2$, $ZnSnAs_2$, $ZnGeP_2$, and ZnO [1]. Recently, Co-doped anatase and rutile $TiO_2$ thin films were reported to be ferromagnetic even above 400 K [2]. These results motivated intensive studies on the structural and electronic properties of these materials. However, due to intrinsic complexities, many questions remain regarding the precise location of Co in the host lattice [3, 4] and the underlying microscopic mechanism of long-range magnetic order.

As far as we know, in the case of rutile $TiO_2$, the theoretical and experimental studies concerned only the Co impurities. In this contribution we present a set of density-functional-theory-based calculations in the systems $R_xTi_{1-x}O_2$ (R = Mn, Fe, Co, Ni) for two impurity concentrations ($x$ = 0.25 and 0.0625). Calculations were performed with the Full-Potential Linearized-Augmented Plane Waves (FLAPW) method assuming that the magnetic impurities substitutionally replace the Ti atoms. Our results show that ferromagnetism appears in all the systems studied, with the exception of $Ni_xTi_{1-x}O_2$. We want to mention that

[*] Corresponding author: e-mail: errico@fisica.unlp.edu.ar, Phone: +54-221-4839061, Fax: +54-221-4252006

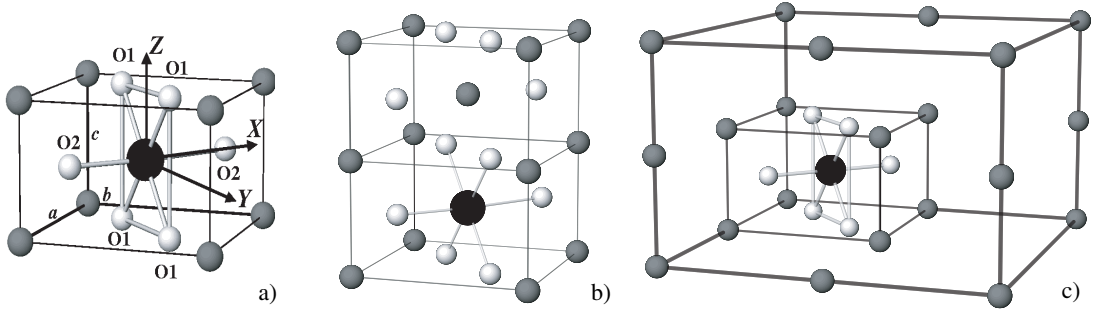

**Fig. 1** a) Unit cell of rutile TiO$_2$ (cations grey balls, O white balls). The results discussed in this work are referred to the indicated axes system, assuming that the impurities replace the black Ti atom. b) 12-atoms SC used in the present work. c) 48-atoms SC used in our calculations. Some O and Ti atoms are not shown for clearness.

some authors claim that Co tends to form clusters in Co$_x$Ti$_{1-x}$O$_2$ [3]. However, the formation of impurity-atom clusters depends not only on the formation energies of different configurations, but also on the microscopic diffusion mechanisms. The determination of the latter would involve an exhaustive investigation of the energetic of impurity complexes, which is not affordable in the present study.

## 2 The system under study. Cell and super-cells

Rutile TiO$_2$ is tetragonal ($a = b = 4.5845_1$ Å, $c = 2.9533_1$ Å [5]). The unit cell (shown in Fig. 1a) contains two metal atoms (Ti) at positions (0, 0, 0) and (1/2, 1/2, 1/2) and four anions (O) at positions $\pm(u, u, 0;$ $1/2 + u, 1/2 - u, 1/2)$ with $u = 0.30493_7$ [5]. The Ti atoms are surrounded by a slightly distorted octahedron with a rectangular basal plane of oxygen atoms O1 with distances to Ti slightly larger than those of atoms O2 of the vertex (see Fig. 1a). To simulate the doped systems R$_x$Ti$_{1-x}$O$_2$ we employed the supercell method. The first super-cell (SC) considered consisted of two unit cells of TiO$_2$ stacked along the $c$-axis with one Ti atom replaced by the magnetic impurity (Fig. 1b). The resulting 12-atoms SC (called 12A-SC in the following) has dimensions $a' = a$, $b' = b$, $c' = 2c$. The impurity concentration in this SC is 25% ($x = 0.25$). The second SC employed consisted of 8 units cells of TiO$_2$ (Fig. 1c), corresponding to an impurity concentration of 6.25% ($x = 0.0625$). The resulting 48-atoms SC (48A-SC) has dimensions $a' = b' = 2a$, $c' = 2c$ and is also tetragonal.

## 3 Method of calculation

The spin-polarized electronic-structure calculations presented in this work were performed with the FLAPW method as embodied in the WIEN97 code [6], in a scalar relativistic version without spin–orbit coupling, which is one of the more accurate schemes for electronic–structure calculations of solids. Exchange and correlation effects were treated within density-functional theory using both the local density (LDA) [7] and the generalized gradient (GGA) [8] approximations. In the FLAPW method the unit cell is divided into non-overlapping ("muffin tin") spheres with radii $R_i$ and an interstitial region. The atomic spheres radii used for Ti and O were 1.01 and 0.85 Å, respectively, while for the magnetic impurities we used $R_i = 1.06$ Å. The parameter $R_{KMAX}$, which controls the size of the basis-set in these calculations, was chosen as 8, that gave 1700 LAPW functions for the 12-atoms SC. For the 48-atoms SC we used $R_{KMAX} = 6$ (3000 LAPW functions). We introduced local orbitals to include Ti-3$s$ and 3$p$, O-2$s$ and Mn-, Fe-, Co-, and Ni-3$p$ orbitals. Integration in reciprocal space was performed using the tetrahedron method taking 200 and 50 $k$-points in the first Brillouin zone for the 12- and 48-atoms SC, respectively. The choice of parameters was checked by performing calculations with increasing number of $k$-points and increasing values of the parameter $R_{KMAX}$.

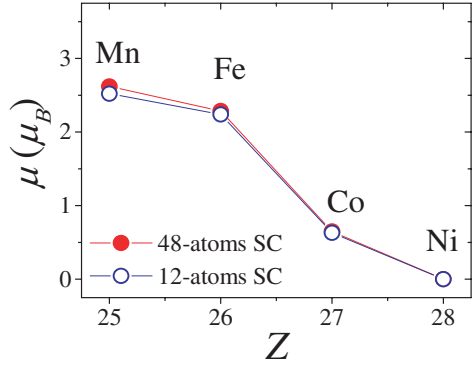

**Fig. 2** Calculated magnetic moments as a function of the atomic number $Z$ of the magnetic impurity for the two impurity dilution studied: 25% (12A-SC) and 6.25% (48A-SC).

## 4 Results and discussion

As the results obtained with the LDA and GGA approximations are very similar, we present here only those obtained using the LDA approximation. In all the cases we found that the presence of the impurities induces local geometrical distortions in the host lattice (essentially in the first oxygen neighbors of the impurities). From the calculated forces acting on the atoms, we found that the impurities tend to reduce the R-O1 and R-O2 bond lengths in about 0.05 Å. In Fig. 2 we present the obtained magnetic moments $\mu$, which are almost independent of the impurity concentration. The results for $\mu$ in the case of $Co_xTi_{1-x}O_2$ are in good agreement with those of Yang et al. [9] for the rutile phase. Park et al. [10], using a different method of calculation, studied Mn, Fe, Co, and Ni in the anatase phase, but their results are not in good agreement with the present ones.

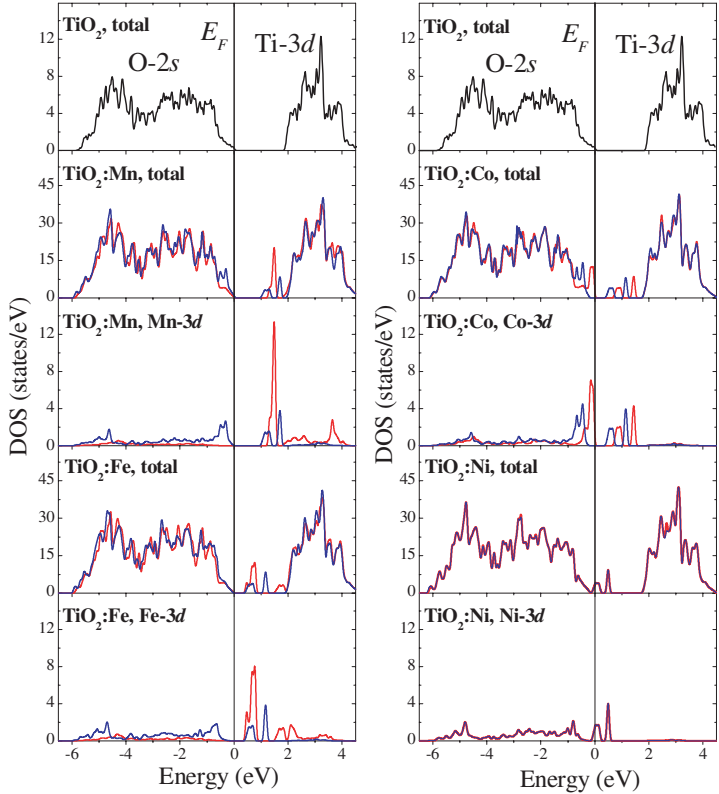

**Fig. 3** Calculated total DOS of the systems $R_xTi_{1-x}O_2$, (R: Mn, Fe, Co, Ni) and 3d-projection at the impurity muffin-tin spheres. In all cases, the DOS correspond to the 48-atoms SCs (impurity dilution 6.25%). The red (blue) lines correspond to the spin-down (spin-up) charges. Energies are refered to the Fermi level.

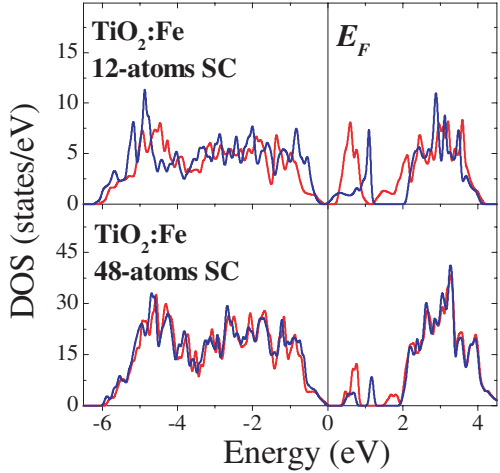

**Fig. 4** DOS of $Fe_xTi_{1-x}O_2$ for $x = 0.25$ and $x = 0.0625$. The red (blue) lines correspond to the spin-down (spin-up) charges. Energies are referred to the Fermi level.

Spin–polarization occurs essentially (for both impurity concentration) at the impurity sites, being the magnetic moment of the other atoms of the SCs nearly zero. Magnetism in $Mn_xTi_{1-x}O_2$ and $Fe_xTi_{1-x}O_2$ is stronger than in $Co_xTi_{1-x}O_2$, and $Ni_xTi_{1-x}O_2$ is not magnetic and $\mu$ exhibits a monotonic behavior with $Z$ (see Fig. 2).

To further study the mechanism of magnetization and exchange coupling, we show in Fig. 3 the density of states (DOS) for pure and doped rutile $TiO_2$ obtained by FLAPW calculations for the 48-atoms SCs. The band-gap for pure rutile $TiO_2$ was found to be about 1.1 eV smaller than the experimental value (3.0 eV [11]) due to the well-known deficiency of LDA.

The local DOS projected in the muffin-tin spheres of the impurities are also shown in Fig. 3. In all cases, the impurity states ($d$ states) are mainly located in the energy-gap of $TiO_2$ and the Ti and O states are not much affected by the doping of the system. The system $Ni_xTi_{1-x}O_2$ is metallic. On the other hand, in $Mn_xTi_{1-x}O_2$, $Fe_xTi_{1-x}O_2$, and $Co_xTi_{1-x}O_2$ the band-gap is drastically reduced, but the insulating electronic structure of the $TiO_2$ host is maintained. We believe the spin interaction between dilute magnetic impurities may be due to super-exchange [12].

All the features discussed for the 48A-SCs are still valid for the 12A-SCs. The only differences are the larger values of the DOS at the Fermi levels and the increased structures from impurity-induced states in the $TiO_2$ gap for the 12A-SCs, pointing to stronger impurity hybridizations in these SCs. As an example we compared in Fig. 4 the DOS of $Fe_xTi_{1-x}O_2$ for the two SCs used. Finally, from a detailed study of the partial DOS projected in the different muffin-tin spheres, we found that the impurity-induced gap states contain $d$ features from the impurities along with contributions from the neighboring O1- and O2-$p$ states. The impurity-states are localized along the $z$ direction but also extend in the $xy$ plane, showing that the oxygen atoms play an important role in mediating magnetic ordering among impurity atoms.

## 5 Conclusions and future studies

In this work we performed a theoretical *ab initio* study of magnetic properties of the systems $R_xTi_{1-x}O_2$ (R: Mn, Fe, Co, Ni) for two impurity concentrations (25% and 6.25%). We found that ferromagnetism appears for all the magnetic impurities, with the exception of Ni. The results presented in this first study show that further more detailed studies are necessary in order to gain a microscopic understanding of the magnetism displayed by these materials. In this sense, we are now extending our theoretical studies to evaluate the possible formation of metallic clusters. From the experimental point of view, EXAFS, SQUID, and Mössbauer experiments (in the case of the system $Fe_xTi_{1-x}O_2$) are in progress. The combination of calculation and experiments will be fundamental to understand not only the microscopic mechanism of long-range magnetic ordering but also the impurity distribution in the $TiO_2$ host.

**Acknowledgements**  This work was partially supported by CONICET, Fundación Antorchas, and ANPCyT (PICT98 03-03727), Argentina. L. A. E. is fellow of CONICET. M. W. and M. R. are members of CONICET. We wish to thank Dr. F. Sánchez, Dra. C. Rodríguez Torres, and Dra. F. Cabrera for fruitful discussions.

## References

[1] H. Ohno, Science **281**, 951 (1998).
    S. B. Ogale et al., Phys. Rev. Lett. **91**, 77205 (2003).
    G. A. Medvedkin et al., J. Appl. Phys. Part 2 **39**, L949 (2000).
    S. Choi et al., Solid State Commun. **122**, 165 (2002).
    S. Cho et al., Phys. Rev. Lett. **88**, 257203 (2002).
    P. Sharma et al., Nature Mater. **2**, 673 (2003).
[2] Y. Matsumoto et al., Science **291**, 854 (2001).
[3] J. Y. Kim et al., Phys. Rev. Lett. **90**, 17401 (2003).
    P. A. Stampe et al., J. Appl. Phys. **92**, 7114 (2002).
    S. A. Chambers et al., Appl. Phys. Lett. **82**, 1257 (2003).
[4] Y. L. Soo et al., Appl. Phys. Lett. **81**, 655 (2002).
[5] J. Hill and C. J. Howard, J. Appl. Cryst. **20**, 467 (1987).
[6] P. Blaha, K. Schwarz, and J. Luitz, Wien 97 (K. Schwarz, TU Wien, Austria, 1999), ISBN 3- 9501031-0-4.
[7] J. P. Perdew and Y. Wang, Phys. Rev. B **45**, 13244 (1992).
[8] J. P. Perdew, K. Burke, and M. Ernzerhof, Phys. Rev. Lett. **77**, 3865 (1996).
[9] Z. Yang, G. Liu, and R. Wu, Phys. Rev. B **67**, R60402 (2003).
[10] M. S. Park, S. K. Kwon, and B. I. Min, Phys. Rev. B **65**, R161201 (2002).
[11] J. Pascual, J. Camassel, and H. Mathieu, Phys. Rev. B **18**, 5606 (1978).
[12] N. W. Ashcroft and N. D. Mermin, Solid State Physics (Harcourt, Orlando, FL, 1995), p. 681.